\documentclass{aastex}
\usepackage{spr-astr-addons}
\shorttitle{An Interesting Paper}
\shortauthors{Enthusiasticus et al.}
\begin{document}

% LaTeX will automatically break titles if they run longer than
% one line. However, you may use \\ to force a line break if
% you desire. Note the use of ~ on the following line to force a
% space in the running header

\title{Thermal desorption of CH$_{4}$  \\
    ~retained in CO$_{2}$ ice}

% Use \author, \affil, and the \and command to format
% author and affiliation information.
% As in the title, use \\ to force line breaks.

\author{Ramon Luna\altaffilmark{1}}

\author{Carlos Mill\'an\altaffilmark{1}}

\author{Manuel Domingo\altaffilmark{2}}

\author{Miguel \'Angel Satorre\altaffilmark{1}}

\affil{$^1$Escuela Polit\'ecnica Superior de Alcoy (UPV), 03801
Alicante, Spain} \affil{$^2$Escuela Polit\'ecnica Superior de
Ingenier\'ia Geod\'esica, Cartogr\'afica y Topogr\'afica (UPV),
46022 Valencia, Spain}

%\affil{National Optical Astronomy Observatories, Tucson, AZ 85719}

%\affil{RSAA, The Australian National University, Weston Creek, ACT2611, Australia}

% Notice that each of these authors has alternate affiliations, which
% are identified by the \altaffilmark after each name.  Specify alternate
% affiliation information with \altaffiltext, with one command per each
% affiliation.

%\altaffiltext{1}{Visiting Astronomer, Cerro Tololo Inter-American Observatory.
%CTIO is operated by AURA, Inc.\ under contract to the National Science
%Foundation.}
%\altaffiltext{2}{Distinguished Fellow, University of the Dense.}
%\altaffiltext{3}{Present address: Center for Astrophysics,
%    60 Garden Street, Cambridge, MA 02138}
%\altaffiltext{4}{Patron, Vivaldi's Restaurant}

% ___________________________________________
% ABSTRACT
% ___________________________________________
%
% Mark off your abstract in the ``abstract'' environment. In the manuscript
% style, abstract will output a Received/Accepted line after the
% title and affiliation information. No date will appear since the author
% does not have this information. The dates will be filled in by the
% editorial office after submission.

\begin{abstract}
CO$_{2}$ ices are known to exist in different astrophysical
environments. In spite of this, its  physical properties
(structure, density, refractive index) have not been as widely
studied as those of water ice. It would be of great value to study
the adsorption properties of this ice in conditions related to
astrophysical environments. In this paper, we explore the
possibility that CO$_{2}$ traps relevant molecules in
astrophysical environments  at temperatures higher than expected
from their characteristic sublimation point. To fulfil this aim we
have carried out desorption experiments  under High Vacuum
conditions based on a Quartz Crystal Microbalance and additionally
monitored with a Quadrupole Mass Spectrometer. From our results,
the presence of CH$_{4}$ in the solid phase above the sublimation
temperature in some astrophysical scenarios could be explained by
the presence of several retaining mechanisms related to the
structure of CO$_{2}$ ice.
\end{abstract}

% Keywords should appear after the \end{abstract} command. The uncommented
% example has been keyed in ApJ style. See the instructions to authors
% for the journal to which you are submitting your paper to determine
% what keyword punctuation is appropriate.


% ___________________________________________
% BODY OF PAPER
% ___________________________________________
%
% From the front matter, we move on to the body of the paper.
%
% Authors who wish to have the most important objects in their paper
% linked in the electronic edition to a data center may do so by tagging
% their objects with \objectname{} or \object{}.  Each macro takes the
% object name as its required argument. The optional, square-bracket
% argument should be used in cases where the data center identification
% differs from what is to be printed in the paper.  The text appearing
% in curly braces is what will appear in print in the published paper.
% If the object name is recognized by the data centers, it will be linked
% in the electronic edition to the object data available at the data centers
%
% Note that for sources with brackets in their names, e.g. [WEG2004] 14h-090,
% the brackets must be escaped with backslashes when used in the first
% square-bracket argument, for instance, \object[\[WEG2004\] 14h-090]{90}).
%  Otherwise, LaTeX will issue an error.

\section{Introduction}
The physical properties of ices present in astrophysical scenarios
are related to their porous structure. One of them is the ice
capacity to retain molecules above their characteristic
sublimation temperature.

This feature is relevant in the physics of gases present in
atmospheres of some astrophysical environments, for instance:
Massive stars \citep{Viti04}, giant planets \citep{Hersant04} and
objects of the Solar System such as Triton \citep{Rubincam03}.

It is known that water ice is the most abundant molecule in
interstellar icy grain mantles and on some objects of our solar
system, but CO$_{2}$ has been revealed as another abundant
molecule \citep{Graauw96, Gurtler96}, and is even the major
component in some astrophysical scenarios, such as, for instance,
Mars in our Solar System. CO$_{2}$ has been studied from several
points of view: Its optical constants (n, k); integrated
absorption coefficient (A); spectral properties before and after
UV photolysis and ion irradiation, using infrared spectroscopy,
and density have been widely derived in previous works
\citep{Wood82, Hudgins93, Ehrenfreund97, Baratta98}.

In spite of the results referred to above, it is necessary to
continue researching to improve our knowledge of the relationship
between the structure and the adsorption properties of CO$_{2}$ at
low temperatures. This is important because it can influence the
abundance of volatile molecules such as CH$_{4}$, N$_{2}$, CO,
etc.

One of the first experiments performed in this area is that of
\citet{Schulze80}. They studied the density as a function of the
deposition temperature, relating this parameter with the
adsorption capacity of CO$_{2}$ ice, from 4 to 87 K. They found
that the density increases from \mbox{$\sim$ 1.0 g cm$^{-3}$} to
1.7 g cm$^{-3}$ as the deposition temperature increases.  As a
result, the adsorption capacity decreases with the increasing
deposition temperature. They refer to density as the structure
including voids (true density). In this paper, we will use the
same definition.

\begin{figure*}
\begin{center}
\includegraphics[width=120mm]{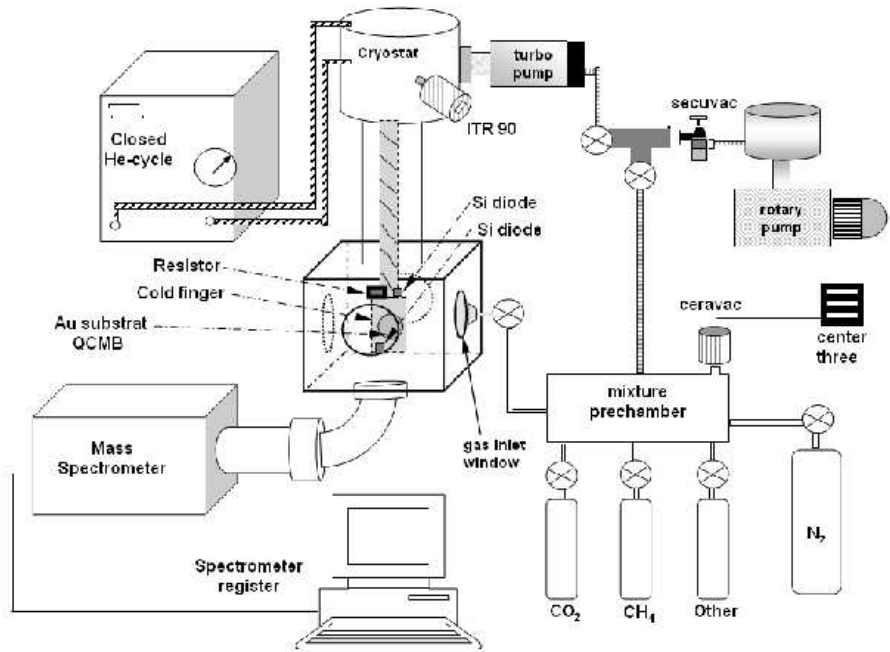}
%\vspace{80mm}
 \caption{Experimental setup.}
 \label{setup}
 \end{center}
\end{figure*}

To the best of our knowledge, there have been no additional
experiments designed to determine the relationship between the
structure and the adsorption capacity of CO$_{2}$ ice. With the
aim of filling this gap, our laboratory is currently carrying out
an exhaustive set of experiments in order to understand the
properties of the CO$_{2}$ structure in different conditions.

The experiments presented in this paper are intended to better
understand the behaviour of relevant astrophysical molecules
co-deposited with CO$_{2}$  and subsequently heated at a fixed
rate. A similar technique, Thermal Programmed Desorption (TPD)
performed under Ultra High Vacuum (UHV) conditions, has been used
to study water  ice structure \citep{Collings03, Collings04}.
Among the structural studies on water it is necessary to highlight
those where the porosity of its amorphous form has been studied
with other techniques \citep{Bar-Nun85, Jenniskens96, Dohnalek03}.
%(com varia estructura d'aigua per difracció d'electrons).
The results and the models used for water by those authors, have
been taken as a starting point despite the fact that the
differences between both molecules are known.

Concerning to water ice, a first result observed is that when the
temperature of deposition increases, the amount of trapped gas
decreases. This is a general trend also found by \citet{Schulze80}
for CO$_{2}$. It is also found that a first release of molecules
is shifted to higher temperatures than their characteristic
sublimation temperature; a second release of molecules is also
recorded at temperatures related with a structural change (phase
change) in water ice ($\sim$ 135 $-$ 150 K); and a last release at
the water sublimation temperature ($\sim$ 160 $-$ 175 K).

All these features are related to the structure of water which is
known to change with temperature. The experiments of
\citet{Stevenson99} and \citet{Jenniskens96} show a continuous
variation of the amorphous structure from 10 K up to the
crystallization temperature to cubic form at around 140 K.

This variation of structure can be seen in the model presented by
\citet{Collings03} based on TPD experiments and RAIR spectra. The
desorption process during crystallization was previously studied
by \citet{smith97} proposing a mechanism known as ``molecular
volcano".

Once the main results have been shown, it is important to
highlight that these experiments are of great importance in many
aspects of astrophysics. An interesting example is that of
\citet{Fraser04}, who showed the relationship between the
mechanisms of chemical and physical adsorption with surface
chemistry under interstellar and protostellar conditions. It could
be also used to model the sublimation of ices as done by
\citet{Viti04} for ices present in solid water near massive
protostars.

The models used by \citet{Viti04} combine codeposition and layered
deposition experiments performed by \citet{Collings04}. In spite
of both kinds of experiments have been used it seems from
comparing the observational data with laboratory experiments that
the water ice shows a compact structure that is best represented
by a layered model rather than a mixed ice \citep{Keane01,
Pontoppidan03, Fraser04b, Guillot04, Palumbo06}. Along the same
lines, this leads to the suggestion that species such as CO,
CO$_{2}$, OCS, CH$_{3}$OH are partially segregated. However, as
\citet{Raut07} have shown, this question could be not completely
clear. Then, the possibility exists that at first the ices were
mixed. With this in mind, our objective is to improve the means at
our disposal with which to improve our knowledge of the chemical
and physical interactions of the observed molecules in space.

Since the existence in different astrophysical environments of
CO$_{2}$ ices is known, it is relevant to study how their
adsorption characteristics can influence the composition of
several environments depending on their physical conditions.

In this paper, the desorption behaviour of CH$_{4}$ from CO$_{2}$
during thermal processing is presented. The experimental setup and
the experimental procedure are explained in Section 2. In Section
3 the main results are shown and discussed. Finally,
%(vore enaltres articles de AA com posa la coma)
the conclusions reached on the influence of temperature on
CO$_{2}$ structure are presented in Section 4.

\section[]{Laboratory experiments}

\subsection[]{Experimental SETUP}
\label{subsection: experiments}

The basic components of our experimental configuration  (Figure
\ref{setup}) to carry out these experiments are a high vacuum and
low temperature system, a quartz crystal microbalance (QCMB), a
laser and a quadrupole mass spectrometer (QMS). The main component
is a high vacuum chamber \mbox{(P $\leq$ 10$^{-7}$ mbar)} whose
pressure conditions are obtained with a rotatory pump ($\sim$
10$^{-3}$ mbar) backing a Leybold TurboVac 50 pump.

The first stage of a closed-cycle He cryostat (40 K) thermally
connected to a shield protector acts as a cryopump providing a
pressure in the chamber below 10$^{-7}$ mbar measured with an ITR
IoniVac transmiter (5 \% in accuracy). The second stage of the
cryostat is named {\emph {cold finger}} and is able to achieve 10
K. Below this, is located the substrate bearing a QCMB (gold
plated surface) in thermal contact with the cold finger.

The temperature in the sample (QCMB) is operated by the
Intelligent Temperature Controller ITC 503S (Oxford Instruments),
using the feedback of a silicon diode sensor (Scientific
Instruments) located just behind it, that lets the temperature
vary between 10 to 300 K with an accuracy of 1 K. Another sensor
is located at the end of the cryostat second stage, on the edge of
the sample holder in order to monitor the behaviour of the system.

Gases or mixtures under study are prepared in a pre-chamber in a
proportion estimated from their partial pressures measured with a
Ceravac CTR 90 (Leybold Vacuum) whose accuracy is 0.2 \%, provided
with a ceramic sensor not influenced by the gas type. The gases
enter the chamber through a needle valve (Leybold D50968) that
regulates the gas flow while the  QMS (AccuQuad RGA 100 with a
resolution of $\sim$ 0.5 amu) allows us to verify the proportion
of gases in the sample (by dividing, in the mass spectra, the area
of methane by the area of carbon dioxide).

\subsection[]{Experimental procedure}

Thermal desorption experiments were carried out to analyze
sublimation temperature of pure and mixed frozen gases. The
following chemicals have been used in this research: CH$_{4}$ $-$
99.9995, CO$_{2}$ $-$ 99.998 (Praxair), N$_{2}$ $-$ 99.999
(Carburos Met\'alicos). Pure gases and mixtures of gases are
prepared for deposition in the pre-chamber. In all the cases the
overall pressure was fixed at 90 mbar.

In order to obtain the desired temperature and to reduce
contamination, the procedure to cool down the cold finger is as
follows: The cryostat is connected and at the same time the
resistor on the cold finger is turned on at a certain voltage to
maintain a temperature in the cold finger (200 K) over the
deposition temperature of undesired gases (mainly H$_{2}$O and
CO$_{2}$). After one hour, when the pressure is around
5$\times$10$^{-8}$ mbar, the current through the resistor is
turned off. This procedure allows us to ensure that only a
negligible amount of contaminants remain in the chamber, taking
into account that a typical experiment lasts 2 hours at maximum.
The deposition temperature (15 K) is achieved in a few minutes (to
again reduce contamination).

Once the temperature is fixed, the needle valve is opened during 1
minute, to fill the chamber with the selected pure gas or mixture
of gases keeping the pumps on. Molecules replenish the chamber
randomly and are deposited onto the QCMB (background deposition).
The amount of deposit is enough to assume that the continuum is
negligible and does not saturate the mass spectrometer (10$^{-4}$
mbar). In all cases the rate of ice deposition is around 1
micrometer per hour, measured using a laser (He-Ne) interferometry
and the QCMB frequency variations.

Once deposited, our experiments were performed by heating the
substrate at a constant rate of 1 K min$^{-1}$, the vacuum system
working continuously, monitoring the molecules present in the
chamber during  desorption with the QMS and checking the molecules
released with the QCMB. The refractive indexes used for
calculations were obtained in our laboratory in a series of
experiments (Satorre et al. in preparation).

\begin{figure*}
\begin{center}
\includegraphics[width=55mm]{lunafg2a.eps2}
\includegraphics[width=55mm]{lunafg2b.eps2}
\includegraphics[width=55mm]{lunafg2c.eps2}
%\vspace{50mm}
 \caption{Frequency variations of QCMB during thermal desorption
  process of gases under our laboratory conditions:
   Left panel: CO$_{2}$; center panel: CH$_{4}$; right panel:
    mixture CO$_{2}$:CH$_{4}$ (95:5).}
 \label{anealing}
 \end{center}
\end{figure*}

\section[]{Results and discussion}

In order to study the capacity of CO$_{2}$ to trap CH$_{4}$ in its
structure we have performed experiments on thermal desorption for
pure CH$_{4}$ and CO$_{2}$ gases  and for mixtures of both them.
%This procedure allowed us to analyse the behaviour
%of pure gases versus mixtures of them.
In the first group of experiments, both molecules were deposited
as a pure film onto the substrate and in the second set, both
gases were co-deposited in a proportion of 95:5
(CO$_{2}$:CH$_{4}$). These three experiments are represented in
Figure \ref{anealing}, where frequency of the QCMB is plotted
versus temperature.

For the thermal process of pure CO$_{2}$ (Figure \ref{anealing},
left panel) we can observe an initial interval (35 to 80 K) where
the frequency varies linearly with the temperature as expected for
our QCMB (in this specific interval of temperatures) when no
release of material takes place. In a second interval, from 80 to
90 K, a sharp increase of frequency is due to the CO$_{2}$
desorption from the QCMB. Finally, at 91 K CO$_{2}$ stops its
desorption and the increase in frequency is again caused by a
linear temperature effect. Hereafter we will take as the
desorption temperature the point in the plot where the slope,
after increasing, changes abruptly. In the case of CH$_{4}$
(Figure \ref{anealing}, center panel), it is shown that the
desorption occurs at 38 K, and further on,  the variation is again
due to temperature effect. In both cases, pure CO$_{2}$ and
CH$_{4}$, only one interval of desorption from the substrate is
observed during the experiment.

Taking into account the desorption temperatures from the QCMB, our
results compare well with those previously published in literature
on this area \citep{Bar-Nun85, Yoshinobu96, Collings03}.

The results of thermal desorption after co-deposition of
CO$_2$:CH$_4$ are presented in Figure \ref{anealing} (right
panel).

\begin{figure}[t]
\begin{center}

\includegraphics[width=80mm]{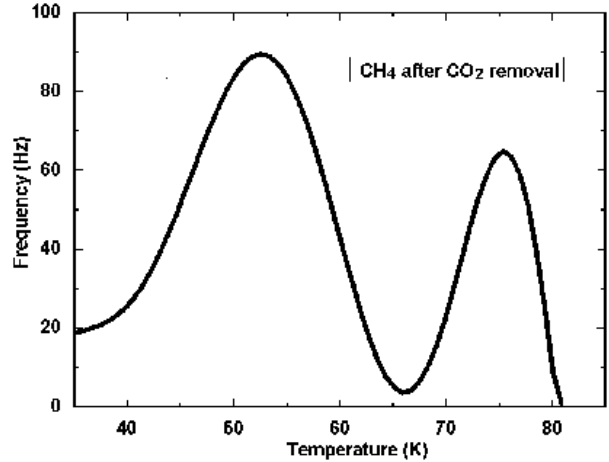}
%\vspace{50mm}
 \caption{CH$_{4}$ remnants detected by QCMB frequency variations after CO$_{2}$ removal and thermal correction.}
 \label{remnants}
 \end{center}
\end{figure}

From the plot, we can observe two features from the frequency
variations that we can associate to CH$_{4}$ mo\-le\-cu\-les
desorbing from the mixture. To isolate the signal due to CH$_{4}$,
we remove the contribution of temperature and CO$_{2}$ release
subtracting the data showed in Figure 2 (left panel), obtaining
Figure \ref{remnants}. In this plot, two previously mentioned
features (the peaks at 50 and 75 K) are clearly visible. Both
signals appear at temperatures higher than the sublimation point
of CH$_{4}$ under our experimental conditions. Therefore, in some
way, CO$_{2}$ has retained CH$_{4}$ molecules.

\begin{figure}[t]
\begin{center}
\includegraphics[width=80mm]{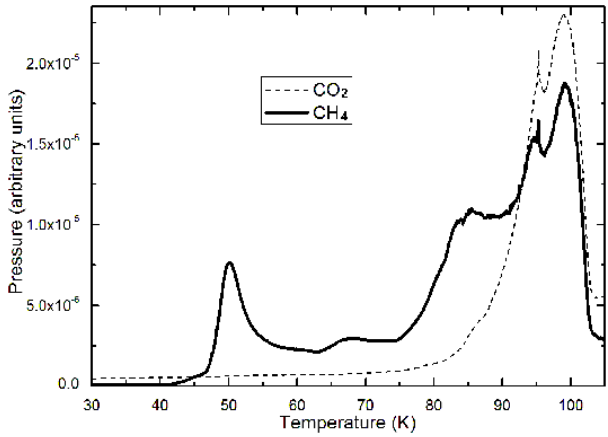}
%\vspace{60mm}
\caption{Thermal desorption process of CO$_{2}$:CH$_{4}$ mixture
recorded as partial pressure versus temperature.}
 \label{em}
 \end{center}
\end{figure}

Since this technique (QCMB) does not allow us to distinguish
different molecules desorbing at the same time, we needed an
additional technique to know whether CH$_{4}$ is retained up to
the characteristic sublimation temperatures of CO$_{2}$ (91 K in
our experimental conditions). Mass spectroscopy allows us to
detect that part of CH$_{4}$ desorbs at the same temperature that
CO$_{2}$ sublimes (Figure \ref{em}). We are able to conclude this
because the behavior of CO$_{2}$ and CH$_{4}$ partial pressure
during thermal desorption from 90 to 110 K are similar.
Additionally, it allowed us to confirm that the peaks at 50 and 75
K are due only to CH$_{4}$ although they appear at slightly
shifted temperatures due to the configuration of our system.

To explain CO$_{2}$ matrix ice, in the literature an amorphous
CO$_{2}$ structure is generally proposed by authors, some of them
arguing that no crystals exist in the whole film when the film is
grown at temperatures below 30 K \citep{Sandford90} but others
suggest that this amorphous structure arises from a compilation of
small crystallites randomly oriented \citep{Schulze80}.

Taking into account our experiments and the previous models just
quoted, below we enumerate and describe the three characteristic
temperatures that we have found with the QCMB and the QMS:

TEMPERATURE 1: Around 50 K, the first release of the gas trapped
by CO$_2$ takes place. This desorption occurs at higher
temperatures than the sublimation point of CH$_4$ and takes place
at the temperature reported previously by other authors as the
transition between amorphous and crystalline phase of solid CO$_2$
\citep{Falk87}. This kind of physical process has been named
molecular volcano by \citep{smith97} in the case of CCl$_4$ in
water.

In the other model, when the temperature increases the
crystallites could undergo a process leading to the structure
compacting from a highly porous one to a less porous structure.
Molecules of CH$_4$ would be linked to the surface of CO$_2$ (we
understand as a surface the upper rough surface and open pores).
The increasing density involves a variation in the characteristics
of the pores, therefore the temperature increase produces the
effect that molecules which were previously retained now sublime.

\begin{figure}[h]
\begin{center}
\includegraphics[width=80mm]{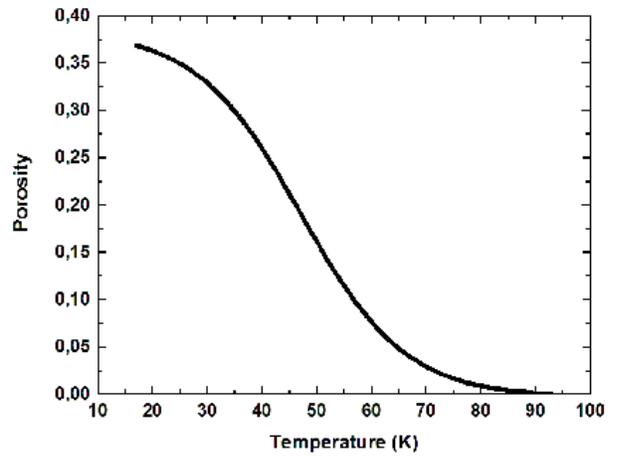}
%\vspace{60mm}
\caption{CO$_{2}$ porosity from Satorre et al. (in preparation).
We have taken our maximum experimental density as the maximum
density in the porosity equation. }
 \label{porosity}
\end{center}
\end{figure}

TEMPERATURE 2: A second release occurs at around 75 K. This fact
coincides with the compacting of the structure that arises from
the continuous variation of the CO$_2$ density, as can be seen
when the porosity reaches the minimum value at around 75 K (Figure
\ref{porosity}). We calculate the porosity as defined by the
equation: $p=1-\frac{\rho_{a}}{\rho_{i}}$, (where $\rho_{a}$ is
the density obtained at certain temperature of deposition and
$\rho_{i}$ is the asymptotic or maximum density), where $\rho_{i}$
= 1.5 g cm$^{-3}$. To perform this plot  the maximum density that
we have taken in this equation is the maximum experimental density
obtained by double angle interferometry and a QCMB when the CO$_2$
is deposited at twelve different temperatures ranging from 15 to
85 K (Satorre et al. in preparation) where the only purpose is to
show how the porosity reaches the minimum value at around 75 K,
not to give a quantitative value of the porosity.

The work of \citet{Schulze80} supports our findings as they show
that the density varies more than 50\% between deposits from 10 K
to 80 K reaching the maximum density at around 75 K.

TEMPERATURE 3: Finally, CH$_4$ molecules are detected when the
CO$_2$ desorbs at 90 K. Those interstitial molecules, which are
the most strongly trapped and have remained inside the structure
after crystallization, are thus retained until desorption of the
CO$_2$ matrix.

\section{Conclusions}

 Our experiments have been used to study the desorption
properties of CH$_4$ in CO$_2$ matrix and to study the structure
of CO$_2$ itself and its interaction with different types of
molecules. These interactions are very complex and requires
further complementary studies (thermal desorption with other
molecules and in different proportions, electron diffraction,...)
considering that the structure of CO$_2$ accreted at low
temperatures is not clear. However some general outlines can be
extrapolated from our results. We found different intervals of
temperature where CH$_4$ is released at higher temperatures than
its sublimation point implying several kinds of interactions
responsible for retaining this molecule within the CO$_2$
structure.

CO$_2$ matrices  could efficiently retain simple mo\-le\-cu\-les.
From the result obtained with CH$_4$ we expect that there are
different mechanisms involved producing various temperatures of
desorption.

A first mechanism could be associated to the beginning of CO$_2$
crystallization with the adsorbed molecules that bring about an
offset to higher temperatures (50 K) than their characteristic
sublimation temperature.

Another kind of interaction can be seen from the onset of the peak
at 70 K corresponding to the  most compacted possible structure of
CO$_2$.

Finally, the molecules more strongly retained in the structure are
revealed from the last sublimation at 90 K.

Once the shifts in the sublimation temperature are described many
astrophysical applications can be found. It is generally assumed
that the volatile components (N$_2$, O$_2$, CO, CH$_4$...) on ice
layers of interstellar dust grains sublime below 40 K. This
assumption may be an oversimplification of the behaviour of such
ices. An important proportion of them may be desorbed into the gas
phase at higher temperatures as a result of adsorption on the
porous surface or entrapment within the closed pores of the
hydrogenated layer (H$_2$O) until it desorbs \citep{Collings04}.
The release of material into the gas phase at higher temperatures
than previously thought may have a significant impact on the
gas-phase chemistry. For example, this has been applied to massive
protostars \citep{Viti04}. But this retarding of the sublimation,
as is evident from the results reported in this paper, would not
be exclusive to the hydrogenated layer (H$_2$O), but would also be
important in the case of the CO$_{2}$ ice. This finding would have
to be taken into account in appropriate scenarios.

\begin{figure}[h]
\begin{center}
\includegraphics[width=80mm]{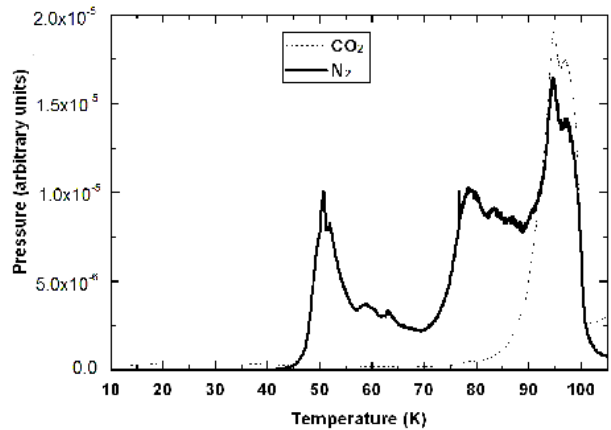}
%\vspace{60mm}
\caption{Thermal desorption process of CO$_{2}$:N$_{2}$ mixture
recorded as partial pressure versus temperature.}
 \label{co2-n2}
 \end{center}
\end{figure}

Furthermore it would be of interest to study the kinetics of
chemical processes at temperatures at which it was previously
thought to be impossible to retain some volatile elements. As
\citet{Fraser04} show, the chemistry is related with the adsorbed
molecules and physical processes (diffusion, adsorption,...) in
the ices.

It is also possible to study the geographical composition of ices
in some solar system satellites such as Triton, where
\citet{Quirico99} explored the possibility that CO$_{2}$ is
segregated in a separated terrain from the other terrains, one
composed of water and the other of a mixture of N$_{2}$, CH$_{4}$
and CO due to difference volatility of molecules. In Triton
CO$_{2}$ may be produced from CO by means of, for example,
chemical reactions with OH radicals as suggested by
\citet{Cruikshank93} or produced by ion irradiation (see for
example \citet{Palumbo93}). In both cases the possibility that
part of the finally segregated CO$_{2}$ retains a small percentage
of volatiles within its structure should not be discarded. In the
light of our results and taking into account that the model of
\citet{Quirico99} fits well but still leaves some discrepancies
with the spectra of Triton, new mixtures with low percentages of
volatiles could be good laboratory candidates to complete the
current models.

Previous applications should be taken as examples, but the
experiments presented here are relevant for any astrophysical
environment in which the presence of CO$_{2}$ ice is important. It
should be taken into account that the conclusions drawn in the
present paper could be applied not only to CH$_{4}$ molecules but
could be enlarged to other simple molecules with similar
characteristics present in astrophysical environments such as
N$_{2}$ whose initial preliminary results (See Figure
\ref{co2-n2}) similar to those of CH$_4$ have corroborated the
shift in temperature although further experiments would be
necessary to confirm this.

\acknowledgments We wish to thank the Ministerio de Educaci\'on y
Ciencia (Co-financed by FEDER funds) AYA2004-05382. We thank the
referee for helpful and constructive comments.


\begin{thebibliography}{}

%\bibitem[Abe \& Shulze, (1979)]{Abe79}  Abe H.,
%Schulze W., 1979, Chemical Physics, 42, 3, 257-263

% The sorption capacity of solid rare gas layer. H. Abe, W.
%Schulze 1979. Chemical Physics, Volumen 42 nº 3. 257-263.

\bibitem[Baratta et al.(1998)]{Baratta98}
Baratta G.~A., Palumbo M.~E., 1998, J. Opt. Soc. Am., 15,
3076-3085

\bibitem[Bar-Nun et al.(1985)]{Bar-Nun85}  Bar-Nun A., Herman G., Laufer D.,
Rappaport M.~L., 1985, Icarus, 63, 317-332

%  Trapping and release of gases by water ice and implications
%for icy bodies. Bar-Nun, A., G. Herman, D. Laufer, and M. L.
%Rappaport 1985. Icarus 63, 317- 332.

%\bibitem[Bar-Nun et al., (1987)]{Bar-Nun87}  Bar-Nun A.,
% Dror J.,  Kochavi E., and  Laufer D., 1987, Phys. Rev. B., 35,

% Amorphous water ice and its ability to trap gases. Bar-Nun, A.,
%J. Dror, E. Kochavi, and D. Laufer 1987. Phys. Rev. B. 35,
%2427-2435.

%\bibitem[Bar-Nun et al., (1988)]{Bar-Nun88}   Bar-Nun A., Kleinfeld I., and  Kochavi E., 1988,
%Phys. Rev., B 38, 7749-7754

%  Trapping of gas mixtures by amorphous water ice. Bar-Nun, A.,
%I. Kleinfeld, and E. Kochavi 1988. Phys. Rev. B 38, 7749-7.


%\bibitem[Cazcarra et al., (1973)]{Cazcarra73}
%Cazcarra V., Bryson II G. E., Levenson L. L., 1973,  J. Vac. Sci.
%Technol. 10, 148

% Spatial distribution of CO$_{2}$ and C2O molecules on reflection
% and sublimation from a cold surface
% V. Cazcarra, G. E. Bryson II and L. L. Levenson,
% J. Vac. Sci. Technol. 10 (1973) 148.


\bibitem[Collings et al.(2003)]{Collings03}
Collings M.~P.,  Dever J.~W., Fraser H.~J., McCoustra M.~R.~S.,
Williams D.~A., 2003, The Astrophysical Journal, 583, 1058-1062

%  Carbon Monoxide entrapment in interstellar ice analogs. M. P.
%Collings et al. The Astrophysical Journal 2003. 583: 1058-1062

\bibitem[Collings et al.(2004)]{Collings04}
Collings M.~P., Anderson M.~A., Chen R., Dever J. W., et al.,
2004, Mon. Not. R. Astron. Soc., 354, 1133-1140

%Collings M.~P., Anderson, M.~A., Chen, R., Dever, J. W., Viti, S.,
%Williams, D.~A., McCoustra, M. R. S., 2004, Mon. Not. R. Astron.
%Soc., 354, 1133-1140

%  A laboratory survey of the themal desorption of
%astrophysically relevant molecules. Mark P. Collings et al. 2004.
%Mon. Not. R. Astron. Soc. 354, 1133-1140.

\bibitem[Cruikshank et al.(1993)]{Cruikshank93}
Cruikshank D.~P., Roush T.~L., Owen T.~C., Geballe T.~R., et al.,
1993, Science, 261, 5122, 742-745



\bibitem[Dohnalek et al.(2003)]{Dohnalek03}
Dohn\'alek Z., Kimmel G.~A., Ayotte T., Smith R.~S., Kay B.~D.,
2003, Journal of Chemical Physics, 118, 1, 364-372

% The deposition angle-dependent density of amorphous solid
%water films, Z. Dohnalek et al., Journal of Chemical Physics
%Volume 118, Number 1 364-372 (2003) 754.


%\bibitem[Ehrenfreund \& Fraser, (2003)]{Ehrenfreund03}
%Ehrenfreund P., Fraser. H., 2003, Science Series II: Mathematics,
%Physics and Chemistry, 120, 317 - 356

% P. Ehrenfreund and H. Fraser, In: Solid state astrochemistry.
%Proceedings of the NATO Advanced Study Institute on Solid State
%Astrochemistry, Erice, Sicily, Italy, 5-15 June 2000, edited by
%Valerio Pirronello, Jacek Krelowski and Giulio Manicò. NATO
%Science Series II: Mathematics, Physics and Chemistry, Vol. 120,
%Dordrecht: Kluwer Academic Publishers, ISBN 1-4020-1558-5, 2003,
%p. 317 - 356

\bibitem[Ehrenfreund et al.(1997)]{Ehrenfreund97}
Ehrenfreund P., Boogert A.~C.~A., Gerakines P.~A., Tielens
A.~G.~G.~M., van Dishoeck  E.~F., 1997, Astronomy and
Astrophysics, 328, 649-669

%Infrared spectroscopy of interstellar apolar ice analogs


\bibitem[Falk(1987)]{Falk87}
Falk M., 1978, Journal of Chemical Physics, 86, 2, 560-564

% Amorphous Solid carbon dioxide

%\bibitem[Fraser et al., (2001)]{Fraser01}
%Fraser H. J., Collings M.P., McCoustra R. S. and Williams D. A.
%MNRAS, 2001, 327, 1165-1172

% Thermal desorption of water ice in the interstellar medium


%\bibitem[Fraser et al., (2002)]{Fraser02}
%Fraser H. J., Collings M.P. and McCoustra R. S., 2002, Review of
%Scientific Instruments, 73, 5, 2161-2170

% Laboratory surface astrophysics experiment



\bibitem[Fraser \& van~Dishoeck(2004)]{Fraser04}
Fraser H.~J., van~Dishoeck E.~F., 2004, Advances in Space
Research, 33, 14-22

% SURFRESIDE: a novel experiment to study surace chemistry under
% interstellar and protostellar conditions

\bibitem[Fraser et al.(2004)]{Fraser04b}
Fraser H.~J., Collings  M.~P.,  Dever J.~W., McCoustra M.~R.~S.,
2004, Mon. Not. R. Astron. Soc., 353, 59-68


\bibitem[de~Graauw et al.(1996)]{Graauw96}
de~Graauw Th., Whittet D.~C.~B., Gerakines P.~A., Bauer O.~H.,  et
al., 1996, Astronomy and Astrophysics, 315, L345-L348

%revisar en ADS



%\bibitem[Grundy et al., (2003)]{Grundy03}
%Grundy W. M., Young L. A., Young E. F., 2003, Icarus, 162, 222

% W. M. Grundy, L. A. Young, E. F. Young, 2003, Icarus, 162,
%222.


\bibitem[Guillot \& Guissani(2004)]{Guillot04}
Guillot B., Guissani Y., 2004, Journal of Chemical Physics, 120,
9, 4366-4382


\bibitem[G\"urtler et al.(1996)]{Gurtler96}
G\"urtler J., Henning T., K\"ompe C., Pfau W., et al., 1996,
Astronomy and Astrophysics, 315, L189-L192

%revisar en ADS


\bibitem[Hersant et al.(2004)]{Hersant04}
Hersant F., Gautier D., Lunine, J.~I., 2004, Planetary and Space
Science, 52, 623-641

% Enrichment in volatiles in the giant planets of the Solar System

\bibitem[Hudgins et al.(1993)]{Hudgins93}
Hudgins D.~M., Sandford S.~A., Allamandola L.~J., Tielens
A.~G.~G.~M., 1993, The Astrophysical Journal S.S., 86, 713-870

%MId- and far- infrared spectroscopy of ices: optical constants and integrated absrobances


\bibitem[Jenniskens \& Blake(1996)]{Jenniskens96}
Jenniskens P., Blake D.~F., 1996, The Astrophysical Journal, 473,
1104-1113


%\bibitem[Kimmel et al., (2001)]{Kimmel01}
%Kimmel G. A., Stevenson, K. P., Dohn\'alek, Z., Smith, R. S.; Kay,
%B. D., 2001, Journal of Chemical Physics, 114, 12, 5284-5294

%   Control of amorphous solid water morphology using molecular
%beams. I. Experimental results. Greg A. Kimmel et al. Journal of
%Chemical Physics  Volume 114, Number 12 5284-5294. (2001)


%\bibitem[Notesco \& Bar-Nun, (1997)]{Notesco97}
%Notesco G., Bar-Nun A., 1997, Icarus, 126, 336-341

% Trapping of methanol, hydrogen cyanide, and n-hexane in water
%ice, above its transformation temperature to the crystalline form.
%G. Notesco, Bar-Nun, A. 1997. Icarus 126, 336-341.

\bibitem[Keane et al.(2001)]{Keane01}
Keane J.~V.,  Boogert A.~C.~A.,  Tielens A.~G.~G.~M., Ehrenfreund
P.,   Schutte W.~A., 2001, Astronomy and Astrophysics, 375,
L43-L46


\bibitem[Palumbo \& Strazzulla(1993)]{Palumbo93}
Palumbo M.~E., Strazzulla, G., 1993, Astronomy and Astrophysics,
269, 568-580

\bibitem[Palumbo(2006)]{Palumbo06}
Palumbo M.~E., 2006, Astronomy and Astrophysics, 453, 903–909


\bibitem[Pontoppidan et al.(2003)]{Pontoppidan03}
Pontoppidan K.~M.,  Fraser H.~J.,  Dartois E.,  Thi W.~F.,
Astronomy and Astrophysics, 2003, 408, 981–1007

\bibitem[Quirico et al.(1999)]{Quirico99}
Quirico E., Dout\'e S., Schmitt B., de~Bergh C., et al., 1999,
Icarus, 139, 159-178


%Quirico E., Dout\'e, S., Schmitt, B., de Bergh, C., Cruikshank, D.
%P., Owen, T.~C., Geballe, T.~R., Roush, T.~L. et al., 1999,
%Icarus, 139, 159-178

% E. Quirico et al., 1999, Icarus, 139, 159.

\bibitem[Raut et al.(2007)]{Raut07}
Raut U., Fam\'a M., Teolis B.~D., Baragiola R.~A., 2007, D.P.S.,
39, 3802R

%Characterization Of Porosity In Vapor-deposited Amorphous Solid Water By Methane Adsorption

\bibitem[Rubincam(2003)]{Rubincam03}
Rubincam D.~P., 2003, Icarus, 163, 2, 469-478

% Polar wander on Triton and Pluto due to volatile migration


\bibitem[Satorre()]{Satorre07}
Satorre et al., in preparation

\bibitem[Sandford \& Allamandola (1990)]{Sandford90}
Sandford S.~A., Allamandola L.~J., 1990, The Astrophysical
Journal, 355, 357-372

% The physuical and infrared spectral properties oc co2 in astrophysical ices analogs


\bibitem[Schulze \& Abe(1980)]{Schulze80}
Schulze W.,  Abe H., 1980, Chemical Physics, 52, 3, 381-388

% Density, refractive index  and sorption capacity of solid CO$_{2}$
%layers. W. Schulze, H. Abe. 1980. Chemical Physics. Volume 52. nº
%3. 381-388.

\bibitem[Smith et al.(1997)]{smith97}
Smith R.~S.,  Huang C., Wong E.~K.~L., Kay B.~D.,  1997, Phys.
Rev. Lett., 79, 909-912

% The Molecular Volcano: Abrupt CCl4 Desorption Driven by the Crystallization of Amorphous Solid Water

\bibitem[Stevenson et al.(1999)]{Stevenson99}
Stevenson K.~P.,  Kimmel G.~A.,  Dohn\'alek Z.,  Smith R.~S., Kay
B.~D., 1999, Science, 283, 1505-1507

% Controlling the Morphology of Amorphous Solid Water. K. P.
%Stevenson et al., Science Vol 283 1505-1507 (1999)


\bibitem[Viti et al.(2004)]{Viti04}
Viti S., Collings M.~P., Dever J.~W., McCoustra M.~R.~S., Williams
D. A.,  2004, Mon. Not. R. Astron. Soc., 354, 1141-1145

% Controlling the Morphology of Amorphous Solid Water. K. P.
%Stevenson et al., Science Vol 283 1505-1507 (1999)

\bibitem[Wood \& Roux(1982)]{Wood82}
Wood B.~E., Roux J.~A., 1982, J. Opt. Soc. Am., 72, 720-728

% Infrared optical properties of thin H20, NH3, and CO2 cryofilms


\bibitem[Yoshinobu \& Kawai(1996)]{Yoshinobu96}
Yoshinobu J., Kawai M., 1996, Surface Science, 368, 247-252

% Absorption and interlayer mixing of methane on Ni (100) at 20 K
%\end{thebibliography}










\end{thebibliography}
\end{document}